\documentstyle[12pt]{article}
\begin{document}
\def\fnote#1#2{
\begingroup\def\thefootnote{#1}\footnote{#2}\addtocounter{footnote}{-1}
\endgroup}
\def\dslash{\not{\hbox{\kern-2pt $\partial$}}}
\def\eslash{\not{\hbox{\kern-2pt $\epsilon$}}}
\def\Dslash{\not{\hbox{\kern-4pt $D$}}}
\def\Aslash{\not{\hbox{\kern-4pt $A$}}}
\def\Qslash{\not{\hbox{\kern-4pt $Q$}}}
\def\Wslash{\not{\hbox{\kern-4pt $W$}}}
\def\pslash{\not{\hbox{\kern-2.3pt $p$}}}
\def\kslash{\not{\hbox{\kern-2.3pt $k$}}}
\def\qslash{\not{\hbox{\kern-2.3pt $q$}}}
\def\np#1{{\sl Nucl.~Phys.~\bf B#1}}
\def\pl#1{{\sl Phys.~Lett.~\bf B#1}}
\def\pr#1{{\sl Phys.~Rev.~\bf D#1}}
\def\prl#1{{\sl Phys.~Rev.~Lett.~\bf #1}}
\def\cpc#1{{\sl Comp.~Phys.~Comm.~\bf #1}}
\def\cmp#1{{\sl Commun.~Math.~Phys.~\bf #1}}
\def\anp#1{{\sl Ann.~Phys.~(NY) \bf #1}}
\def\etal{{\em et al.}}
\def\half{{\textstyle{1\over2}}}
\def\be{\begin{equation}}
\def\ee{\end{equation}}
\def\ba{\begin{array}}
\def\ea{\end{array}}
\def\tr{{\rm tr}}
\def\Tr{{\rm Tr}}
\title{Quantization of a self-interacting maximally charged string
\thanks{Research
supported by the DoE under grant DE--FG05--91ER40627.}}
\author{\sc
Suphot Musiri\fnote{\dagger}{E-mail: \tt smusiri@utk.edu} and George Siopsis
\fnote{\ddagger}{E-mail: \tt gsiopsis@utk.edu}\\
\em Department of Physics and Astronomy, \\
\em The University of Tennessee, Knoxville, TN 37996--1200, USA.\\
}
\date{August 2001}
\maketitle
\vskip -3.0in\hfill\hfil{\large UTHET--01--0802}\vskip 250pt
\begin{abstract}
We discuss the quantization of a self-interacting string consisting of maximally charged matter.
We construct the Hamiltonian in the non-relativistic limit by expanding around
a static solution of the Einstein-Maxwell field equations.
Conformal symmetry is broken on the worldsheet,
but a subgroup of the conformal group acts as the gauge
group of the theory. Thus,
the Faddeev-Popov quantization procedure of fixing the gauge is applicable.
We calculate the Hamiltonian and show that, if properly quantized, the system possesses a
well-defined ground state and the spacing of its energy levels is of order the Planck mass.
This generalizes earlier
results on a system of maximally charged black holes to the case of continuous matter distributions.
\end{abstract}
\renewcommand\thepage{}\newpage\pagenumbering{arabic} 

The dynamics of gravitating matter distributions is in general a hard problem to solve. In the extremal
case of maximally charged matter, this problem can be studied by starting with a static solution of
the field (Einstein-Maxwell) equations and then quantizing the fluctuations around such solutions using
perturbation theory. This was first discussed by Ferrell and Eardley~\cite{bib5}. It was subsequently
realized that for a discrete matter distribution there was an enhanced conformal symmetry when
the matter particles (black holes) were close together~\cite{bib4}. This created problems for the
quantization of the system. In general, 
systems with conformal symmetry are problematic quantum systems being described by
Hamiltonians with no well-defined ground state. A solution to this problem was suggested by
de Alfaro, Fubini and Furlan
(DFF)~\cite{bib1}. They proposed the
redefinition of the Hamiltonian by the addition of a potential term proportional to
the generator of special conformal transformations.

In the two-particle (black hole) case, the DFF redefinition of the Hamiltonian can be interpreted as
a redefinition of the time coordinate. The DFF Hamiltonian corresponds to a
globally defined time coordinate whereas the conformally invariant definition
does not. Thus, the DFF trick appears plausible on physical grounds~\cite{bib2,bib2a,bib2b,bib2c,bib3}.
Unfortunately, this physical picture does not straightforwardly generalize to the case of
multiple black hole scattering~\cite{bib4}.

In ref.~\cite{bib7}, an alternative justification of the DFF
procedure was introduced for a multi-particle (black hole) system.
It was noted that the system possessed a gauge invariance due to the reparametrization
invariance of the description of the particle (black hole) orbits.
It was shown that the redefinition of the Hamiltonian amounted to a
different choice of gauge. In the conformally invariant case, an
obstruction to the standard gauge-fixing procedure was identified that led to a
modification of the usual quantization rules. This obstruction came from
the boundary of moduli space and was rooted in the fact that the time coordinate
was not defined at the boundary. On the other hand, there was no obstruction
in the choice of gauge leading to the DFF Hamiltonian. It was concluded that the
DFF Hamiltonian corresponded to a good gauge choice, whereas the conformally
invariant Hamiltonian did not. The discussion was based on the standard
Faddeev-Popov quantization procedure and was therefore applicable to more
general systems, as long as the system had an underlying gauge invariance.

Here, we extend the procedure discussed in~\cite{bib7} to the case of
a continuous one-dimensional matter distribution (string).
The string is charged, so worldsheet conformal invariance is broken.
However, the system still possesses reparametrization invariance of the two parameters
of the worldsheet independently of each other. This is a gauge symmetry and needs to
be fixed when quantizing the system.
We will show how the gauge can be fixed
without encountering obstructions from the singularities of
moduli space. The resultant Hamiltonian contains a
potential term, as prescribed by the DFF trick. Thus, we show that the DFF trick is a
consequence of a standard gauge-fixing procedure in the case of a continuous one-dimensional
matter distribution (string). 

Concentrating on five spacetime dimensions and choosing units so that the Planck mass is
$M_{Pl} = 1$, the action may be written as
\be
S = S_{fields} + S_{matter}
\ee
where the action for the fields is
\be
S_{fields} = {1\over 12\pi^2} \int d^5 x\, \sqrt{-g} \Big(R - {\textstyle{3\over
4}} F^2\Big) + {1\over 12\pi^2} \int A \wedge F\wedge F
\ee
in terms of a dynamical metric field $g_{\mu\nu}$ and electromagnetic vector
potential $A_\mu$, both functions of the coordinates $x^\mu$ ($\mu = 
0,1,\dots,4$).
Matter is described by the mass (charge) density $\rho$ and the
current $j^\mu = \rho v^\mu$, where $v^\mu = \partial X^\mu / \partial\tau$ is the four-velocity.
We are interested in a one-dimensional continuous matter distribution. Then the position vector
$X^\mu$ spans a two-dimensional surface, $X^\mu (\sigma, \tau)$, and the action is
\be\label{eqsmatter}
S_{matter} = \int d \sigma d\tau\, \sqrt{-g} \, A_\mu j^\mu - \int d \sigma d\tau\, \sqrt{-g} \, \rho
\ee
where we integrate over the worldsheet of the string,
together with the constraint
\be\label{eq62}
\nabla_\mu j^\mu = 0
\ee
which enforces local conservation of charge and ensures gauge invariance.
Notice that if the density $\rho$ is constant and the vector potential is absent ($A_\mu = 0$), the
action~(\ref{eqsmatter}) reduces to the Nambu-Goto action for a string that possesses conformal
invariance. In our case, conformal invariance is broken, but reparametrization of $\tau$ and $\sigma$
separately is still a symmetry of the theory. This is a gauge symmetry and we need to fix the
gauge. The naive gauge choice would be $X^0 = \tau$. However, it was shown in ref.~\cite{bib7} that in the discrete case, there is a subtlety in the application of the Faddeev-Popov procedure if one
adopts the gauge $X^0 = \tau$ due to contributions from the boundary of moduli space. Thus, the choice
$X^0 = \tau$ was not a good gauge-fixing condition and a different gauge choice had to be made that
was free of obstructions from the boundary of moduli space. In the continuum case considered here,
similar obstructions are present. This is expected, because the continuum case can be viewed as a limiting case of a discrete distribution of matter. Therefore, we need to be careful in applying the quantization
procedure.

The Einstein-Maxwell equations admit static solutions, for which $\vec j = \vec 0$. We will first study such a solution and then perturb around the static configuration. A static one-dimensional configuration
is given by
\be\label{eqle}
ds^2 = - {1\over \psi^2}\, dt^2 + \psi\, d\vec 
x^2
\quad,\quad
A = {1\over\psi}\, dt 
\ee
where (in flat space)
\be
\vec\partial^2 \psi = - 4\pi^2 \psi^2 \rho
\ee
In terms of the modified charge (mass) density
\be
\widetilde\rho = \psi^2 \, \rho
\ee
we obtain
\be
\psi (\vec x) = \int d \sigma \sqrt{g} \; {\widetilde\rho [\vec X(\sigma)]\over (\vec x - \vec X(\sigma))^2}
\ee
where $g = (\partial \vec X (\sigma) /\partial\sigma )^2$, which is the Coulomb potential due to a line charge distribution in four spatial dimensions. Since we are interested in describing the self-interactions
of this string, we need to be able to take the limit as $\vec x$ approaches the string $\vec X(\sigma)$.
To safely do so, we introduce a small dimensionless parameter $\delta$ (UV cutoff)
and the regulated potential
\be\label{eqreg}
\psi_{reg} (\vec x) = \int d \sigma \sqrt{g} \; {\widetilde\rho [\vec X(\sigma)]\over (\vec x - \vec X(\sigma))^2 + L^2\delta^2 }
\ee
where $L$ is the physical length of the string.
Now consider the motion of a small segment of the string of length $2L\delta$ under the influence of
the rest of the string. If the segment is at $\sigma = \bar\sigma$, then the potential it experiences is
\be
\bar\psi = \psi_{reg} (\vec X (\bar\sigma)) = \int d \sigma \sqrt{g} \; {\widetilde\rho [\vec X(\sigma)]\over (\vec X(\bar\sigma) - \vec X(\sigma))^2 + L^2\delta^2 }
\ee
The leading contribution to this integral comes from the neighbourhood of $\bar\sigma$ where
$\vec X (\sigma) \approx \vec X (\bar\sigma) + (\sigma - \bar\sigma) \partial\vec X/\partial\sigma$.
If $\widetilde\rho$ is a sufficiently slowly varying function, we obtain
\be
\bar\psi = \int dl \; {\bar\rho \over l^2 + L^2\delta^2} + \dots = {2\bar\rho\over L\delta} + \dots = {m\over L^2\delta^2} + \dots
\ee
where $\bar\rho = \widetilde\rho [\vec X(\bar\sigma)]$ is the density at $\bar\sigma$, $m=
2\bar\rho L\delta$ is the mass of the segment, $l = \sqrt g\, |\sigma -\bar\sigma|$ is the distance along the string measured from $\bar\sigma$
and the dots represent higher-order terms in $\delta$.
If we place the origin at $\vec X(\bar\sigma)$ and approximate the segment by a point particle of mass
$m$ at distance $r = L\delta$, we can have radial motion under which the length of the segment
will change and angular
motion leaving its length unchanged. The line element along its trajectory can then be written as~({\em cf.}~eq.~(\ref{eqle}))
\be dS^2 = - {1\over \bar\psi^2} (dX^0)^2 + \bar\psi (d\vec X)^2 = - {L^2\delta^2\over m} (dX^0)^2
+ {m\over L^2} dL^2 + m \, d\Omega_3^2\ee
which is a metric in $AdS_2\times S^3$. This line element may also be written as
\be dS^2 = - {1\over\bar\psi^2} \left( (dX^0)^2 - {m\over 4}\, d\bar\psi^2 \right) + m\, d\Omega_3^2\ee
which describes the motion of a particle in the vicinity of a Reissner-Nordstr\"om black hole of mass
$m$, provided $\bar\psi \gg 1$. This requirement translates to $m\gg L^2\delta^2$, i.e., $\bar\rho\gg
L\delta$, which is certainly satisfied since $\bar\rho /L \sim o(1)$.

We may now apply the results of ref.~\cite{bib7} to study the dynamics of the string segment.
It is convenient to introduce coordinates
\be x^\pm = X^0 \pm {\sqrt m\over 2} \, \bar\psi \ee
and conjugate momenta $p_\pm$. The generator of gauge transformations ($\tau$ reparametrizations)
may be written as
\be 2m \chi = -\bar\psi^2 p_+p_- + \half m\bar\psi (p_++p_-) + {L^2\over m} = 0\ee
where $\vec L$ is the angular momentum operator. The system is constrained by
\be \chi = 0\ee
There are three conserved quantities (that commute with $\chi$)
\be h = - p_+-p_-\quad,\quad d = 2x^+p_++2x^-p_-\quad,\quad k = -(x^+)^2p_+ -(x^-)^2p_- + \half
m^2\bar\psi \ee
generating time ($X^0$) translations, dilatations and special conformal transformations, respectively.
These three quantites obey an $SL(2,{\mathbf R})$ algebra
\be \{\, h\;,\; d\, \} = -2h\quad,\quad \{\, h\;,\; k\, \} = -d\quad,\quad \{\, k\;,\; d\, \} = 2k\ee
reflecting the symmetry of the $AdS_2$ spacetime. The brackets may be Poisson or Dirac, so this is
also an algebra of the gauge-fixed system, as expected.

The simplest gauge-fixing condition
\be\label{eqgf1} X^0 = \tau\ee
leads to a system with Hamiltonian $h$. In the non-relativistic limit, the Hamiltonian becomes
\be h \approx {2\bar\psi p^2\over m^2} \ee
where $p$ is the momentum conjugate to $\bar\psi$.
The action is
\be s = \int d\tau \; \left( p \, {d\bar\psi\over d\tau} - h \right) \ee
After integrating over the momentum in the path integral, this takes the form
\be s = \int d\tau \; \half m \bar\psi^2 v^2 \ee
where $v = \dot r$  is the velocity of the center of mass of the segment ($r = L\delta$).
Summing over all the segments of the string, the total action~(\ref{eqsmatter}) in the non-relativistic limit becomes
\be\label{eqsmatter2} S_{matter} = \half \int d\tau dl \widetilde\rho \psi^2 v^2 \ee
for the static configuration~(\ref{eqle}). Notice that this form of the action can also be directly
derived by taking the non-relativistic limit of eq.~(\ref{eqsmatter}).

The system described by~(\ref{eqsmatter2}) does
not have a well-defined vacuum. The origin of the problem was traced to an obstruction in the gauge-fixing procedure due
to a boundary contribution to the path integral~\cite{bib7}. Thus the naive gauge $X^0 = \tau$ is not
a good gauge. A set of good gauges (free of boundary contributions) is given by the gauge choice
\be\label{eqgf2} \tau (x^+,x^-) = \arctan \left( {\omega x^+ + \omega x^-\over 1-\omega^2 x^+x^-} \right) = \tau\ee
The conjugate momentum to $\tau$ is
\be p_\tau = -\half \left( {p_+\over \partial_+ \tau} + {p_-\over \partial_- \tau} \right) = {1\over 2\omega}\,
(h+\omega^2 k') \quad,\quad k' = -(x^+)^2p_+ -(x^-)^2p_-\ee
This is not a conserved quantity ($\{\, h'\,,\, \chi\, \} \ne 0$). To remedy this, we perform a gauge
transformation on the vector potential, $A\to A+d\Lambda$, whose effect is the addition of a total
time derivative to the action~(\ref{eqsmatter}). The choice
\be \Lambda = - {m^{3/2}\over 4}\, \ln {1+\omega^2 (x^+)^2\over 1+\omega^2 (x^-)^2} \ee
leads to the new conjugate momentum
\be h' = p_\tau -\partial_\tau\Lambda
= {1\over 2\omega}\,
(h+\omega^2 k) \ee
which is a conserved quantity (since both $h$ and $k$ are).
$h'$ is the Hamiltonian of the system after gauge-fixing and has a well-defined vacuum. In the non-relativistic limit, we obtain
\be h' \approx {1\over 2\omega}\, \left( {2\bar\psi p^2\over m^2} + \half m^2\omega^2\bar\psi \right)\ee
where $p$ is the momentum conjugate to $\bar\psi$. This is the Hamiltonian of a harmonic oscillator, as
can be seen by performing the transformation $\bar\psi = C x^2$. The energy levels are all equally
spaced and the spacing is $o( M_{Pl})$. This also shows that the energy levels are independent of
$\omega$ which is an arbitrary variable (different values of $\omega$ are related to each other
through gauge transformations).

The action in the Lagrangian picture can be found as before (see derivation of eq.~(\ref{eqsmatter2})),
\be\label{eqsmatter3} S_{matter} = \int dt dl\, \left( \half \widetilde\rho \psi^2 v^2 - \widetilde\rho^3 \right)\ee
where $t = X^0$ and $\vec v = \partial \vec X /\partial X^0$.
Notice that the gauge parameter $\omega$ has disappeared, as expected.

To obtain non-static configurations, we shall perturb around the static
solution in the non-relativistic limit.
Using the {\em ansatz}
\be
ds^2 \equiv g_{\mu\nu}\, dx^\mu dx^\nu = - {1\over \psi^2}\, dt^2 + \psi\, d\vec 
x^2
+ 2\vec N \cdot d\vec x\, dt
\ee
\be
A = {1\over\psi}\, dt + \vec A \cdot d\vec x
\ee
and keeping only terms quadratic in the potentials $\vec N,\vec A$ and 
discarding total derivatives,
the action for the fields becomes~\cite{bib4}
$$S_{fields} = {1\over 12\pi^2} \int d^5 x \; \left( 3\partial_t
\vec P\cdot\vec\partial \psi - {3\over 4\psi}\, F^2 + {3\over 2\psi^2}\; FG -
{1\over 2\psi^3} \; G^2 \right. $$
\be
\left. - 3\psi (\partial_t\psi)^2 - {3\lambda\over 4\psi} \; F\widetilde F +
{3\lambda\over 4\psi^2} \; F\widetilde G - {\lambda\over 4\psi^3} \; G\widetilde G\right)
\ee
where $\lambda = 1$ in the supersymmetric case.
We introduced the convenient (gauge-invariant) combinations
\be
\vec P = \vec A + \psi \vec N\quad,\quad \vec R = \psi^2 \vec N
\ee
whose field strengths respectively are
\be
F_{ij} = \partial_i P_j - \partial_j P_i\quad,\quad G_{ij} = \partial_i R_j
-\partial_j R_i
\ee
and their duals: $\widetilde F^{ij} = \epsilon^{ijkl} F_{kl}$,
$\widetilde G^{ij} = \epsilon^{ijkl} G_{kl}$.

The action for the matter can be manipulated as in the static case. The effect of the perturbation
is the addition of a term linear in the potential $\vec P$.
We obtain
\be\label{eqsmatter4}
S_{matter} = \int dtdl \; \widetilde\rho (\half\psi^2\vec v^2 - \widetilde\rho^2 + \vec v\cdot\vec P)
\ee
({\em cf.}~eq.~(\ref{eqsmatter3})) where we adopted the gauge-fixing condition~(\ref{eqgf2}).
Had we adopted the gauge-fixing condition~(\ref{eqgf1}) instead, we would have obtained an
action with the middle term absent, describing a system without a well-defined vacuum.
The action~(\ref{eqsmatter4}) consists of two pieces; one is a potential term independent of the
velocities, the other is a kinetic term quadratic in the velocities. To emphasize this, we will write
the action as
\be
S_{matter} = \int dt \; (T_{matter} - V)\ee
where
\be\label{eqtv} T_{matter} = \int dl\, \widetilde\rho (\half\psi^2\vec v^2 + \vec v\cdot\vec P) \quad,\quad
V = \int dl\, \widetilde\rho^3\ee
The equations of motion are the Einstein and Maxwell Equations which yield
\be
\psi = {\cal A}_0\quad,\quad F_{ij} = 2\psi\, {\cal F}_{ij}\quad,\quad
G_{ij} = 3\psi^2\, {\cal F}_{ij}
\ee
where ${\cal A}_\mu$ is the vector potential generated by the source current
$\psi^2 j^\mu$ in flat spacetime and ${\cal F}_{\mu\nu}$ is its field strength:
\be
\partial_\mu {\cal F}^{\mu\nu} = 2\pi^2 \psi^2 j^\nu \quad,\quad
{\cal F}_{\mu\nu} = \partial_\mu {\cal A}_\nu - \partial_\nu {\cal A}_\mu
\ee
In the non-relativisic limit ($|\vec v|\ll 1$),
the vector potential is found to be
\be
{\cal A}_0 (t, \vec x)= \psi (t, \vec x)= \int d\sigma\, \sqrt g \, \Delta (\vec x, \vec X (t,\sigma)) \widetilde\rho (t,\sigma) \quad,\quad
\vec{\cal A} (\vec x)= \int d\sigma\, \sqrt g \, \Delta (\vec x, \vec X (t,\sigma) \widetilde\rho (t,\sigma))
\vec v (\sigma)
\ee
where $\widetilde\rho (t,\sigma)$ is shorthand for $\widetilde\rho [t, \vec X (t,\sigma)]$ and
\be
\Delta (\vec x,\vec y) = {1\over (\vec x - \vec y) + \delta^2} \ee
is the regulated propagator ({\em cf.}~eq.~(\ref{eqreg})).
This is the Lorentz gauge solution,
\be
\partial_\mu {\cal A}^\mu = -\partial_t\psi + \vec\partial\cdot\vec{\cal A} =0
\ee
Notice that there are two terms in the action that are given in terms of the
potential $\vec P$
(all other terms involve the field strengths and $\psi$ only),
\be
12\pi^2 \widetilde\rho\vec v\cdot\vec P + 3\partial_t
\vec P\cdot\vec\partial \psi 
\ee
They can be manipulated as follows. First observe that
\be
\partial_i\partial_t\psi = \partial_i \vec\partial\cdot\vec{\cal A} =
\partial_j {\cal F}_{ij} + \vec\partial^2 {\cal A}_i = \partial_j {\cal F}_{ij} + 4\pi^2\widetilde\rho
\vec v
\ee
where in the last step we used the Maxwell equations.
Therefore, up to total derivatives,
\be
12\pi^2 \widetilde\rho\vec v\cdot\vec P + 3\partial_t
\vec P\cdot\vec\partial \psi = 3P_i \partial_j {\cal F}_{ij}
= {\textstyle{3\over 2}} F_{ij} {\cal F}_{ij}
= 3\psi {\cal F}^2
\ee
where we only include the magnetic field in ${\cal F}^2$.
The term in the action that involves a time derivative can be similarly
manipulated. After some algebra, we obtain
\be
-3\psi (\partial_t\psi)^2 = 3{\cal A}_i {\cal F}^{ij}\partial_j\psi
+3\psi {\cal F}^2 -3\psi\vec\partial^2\psi \vec{\cal A}\cdot\vec v
\ee
Therefore, up to total derivatives, the kinetic piece of the action can be written as
\be
S_{kinetic} = S_{fields} + \int dt\, T_{matter} = {1\over 4\pi^2} \int d^5x \; \left(
\psi \vec\partial^2\psi (\psi\vec v^2 + \half \vec{\cal A}\cdot\vec v)
+ {\cal A}_i {\cal F}^{ij} \partial_j
\psi
+ \psi {\cal F}^2 
-{\lambda\over 4}\, \psi {\cal F} \widetilde{\cal F} \right)
\ee
The total action is the sum of the kinetic piece $S_{kinetic}$, which is quadratic in the
velocity field, and the potential term, $\int dt\, V$~(see eq.~(\ref{eqtv})).
To show the explicit dependence of the kinetic piece of the action on the velocity field, we write
\be
S_{kinetic} = \int dt\, T\ee
\be\label{eqkin} T ={1\over 12\pi^2} \int dl\, dl'\, dl'' \;
{\cal T} [\vec v (\sigma), \vec v(\sigma') ; \vec X(\sigma), \vec X(\sigma'), \vec X(\sigma'')]\widetilde\rho (t,\sigma)
\widetilde\rho (t,\sigma')\widetilde\rho (t,\sigma'')
\ee
where $dl = d\sigma \sqrt g$ ($g = (\partial\vec X/\partial\sigma)^2$), and similarly for $dl'$ and $dl''$,
are line elements along the string, and we introduced the kernel
\be\label{eqtij}
{\cal T} [\vec v_1, \vec v_2; \vec x,\vec y,
\vec z] = \int d^4 u \Big( (\vec v_1^2 - \vec v_1\cdot\vec v_2)\, \vec\partial^x\cdot\vec\partial^y
+v_1^iv_2^j (\partial_{[i}^x\partial_{j]}^y + \lambda
\epsilon_{ij}^{\;\;\;\; kl} \partial_k^x\partial_l^y ) \Big)\Delta (\vec u ,\vec x)
\Delta (\vec u ,\vec y)\Delta (\vec u ,\vec z)
\ee
To compute ${\cal T}$, it is convenient to apply the derivatives after we perform the integration. Introducing Feynman parameters, we obtain
\be
\int d^4 u \Delta (\vec u ,\vec x)
\Delta (\vec u ,\vec y)\Delta (\vec u ,\vec z) = \pi^2 \int [d\alpha] {1\over
D}
\ee
where
$[d\alpha] = d\alpha_1d\alpha_2d\alpha_3 \; \delta (1-\alpha_1-\alpha_2-\alpha_3)$ and
\be
D =
(\vec x - \vec y)^2 \alpha_1\alpha_2 + (\vec y - \vec z)^2 \alpha_2\alpha_3 +(\vec z - \vec x)^2 \alpha_3\alpha_1 +\delta^2
\ee
The three-point function ${\cal T}$ is finite. The apparent logarithmic
singularities are canceled by the derivatives. The total action is
\be S = \int dt\, (T-V) \ee
where $V$ is the potential term~(\ref{eqtv}). The presence of the potential term ensures the existence
of a ground state. The spacing of the energy levels of the Hamiltonian derived from this action is $o(M_{Pl})$.

As an example consider the case of a circle of radius $R$. To find the Hamiltonian for the modulus $R$,
further assume that matter is uniformly distributed along the string and the velocity field is radial and uniform, so that the system is described by a single
modulus, $R(t)$. The total mass of the string is
\be\label{eqch} M = 2\pi R \widetilde\rho \ee
Since $M$ is a constant, the density $\widetilde\rho$ will be changing in time according to~(\ref{eqch}).

We start by computing the kinetic energy $T$~(\ref{eqkin}). A short calculation yields
\be T = {M^3\over 8\pi^5} \int d\sigma d\sigma' d\sigma'' \; {\cal T}\; \dot R^2\ee
Using~(\ref{eqtij}) and integrating over the string parameters (angles) $\sigma, \sigma', \sigma''$,
we obtain
\be
T = {1\over 4\pi^2} M^3 {\dot R^2\over R^4}
\ee
The potential $V$ (eq.~(\ref{eqtv})) is
\be V = R\int d\sigma \widetilde\rho^3 = {M^3\over 4\pi^2R^2} \ee
Therefore, the Lagrangian is
\be L = T - V = {1\over 4\pi^2} M^3 {\dot R^2\over R^4} - {M^3\over 4\pi^2R^2}
\ee
The Hamiltonian is
\be
H = {4\pi^2 R^4P^2\over M^3} + {M^3\over 4\pi^2R^2}
\ee
where $P$ is the momentum conjugate to $R$.
This
is equivalent to a harmonic oscillator Hamiltonian. To see this, change variables to
\be
u = {1\over\sqrt\pi}\, {M\over R}
\ee
If $P_u$ is the conjugate momentum, we have
\be
H = {1\over 2\pi} \left( {P_u^2\over 2M} + \half M u^2\right)
\ee
The spectrum of this Hamiltonian consists of equally spaced energy levels differing by $1/ 2\pi$.
This is in units of the Planck mass $M_{pl}$. Therefore, this approximation is valid only for very
heavy strings ($M \gg M_{pl}$). Notice that otherwise, there is no restriction on the mass $M$,
since this is a solution to the field equations for any $M$, as long as the mass equals the charge
locally everywhere.

In conclusion, we studied the quantization of a self-interacting one-dimensional continous matter
distribution in five space-time dimensions. Matter was maximally charged. We expanded around a
static solution of the field (Einstein-Maxwell) equations and used perturbation theory to quantize
small fluctuations around the static classical solution. The system possessed a gauge invariance
which was a subgroup of the conformal group on the worldsheet of the string. We showed that a
careful application of the Faddeev-Popov procedure produced a Hamiltonian that included a potential
term which ensured the existence of a ground state. We obtained energy levels whose spacing was
$o(M_{Pl})$. This generalized earlier results of ref.~\cite{bib7} for discrete matter distributions.
It would be interesting to move away from the extremal point with an eye toward the limit of
zero charge in which the full conformal symmetry on the worldsheet would be present.

\end{document}